\documentclass{PoS}

\title{Static potential for the quark-antiquark-gluon hybrid system in lattice QCD}

\ShortTitle{Static potential for the quark-antiquark-gluon hybrid system in lattice QCD}

\author{\speaker{Marco Cardoso} \\
         CFTP, Instituto Superior T\'ecnico\\
        E-mail: \email{mjdcc@cftp.ist.utl.pt}}

\author{
Pedro Bicudo \\
CFTP, Instituto Superior T\'ecnico\\
\email{bicudo@ist.utl.pt}
}

\author{
Orlando Oliveira \\
CFC, Universidade de Coimbra\\
\email{orlando@teor.fis.uc.pt}
}

\abstract{
The static gluon-quark-antiquark interaction is investigated using lattice QCD techniques. A Wilson loop adequate to the static hybrid three-body system is developed and, using a $24^3 \times 48$ periodic lattice with $\beta = 6.2$, the potential energy of the system is measured for different geometries. For the medium range behaviour, when the quarks are far apart, we find a string tension which is compatible with two fundamental strings. On the other hand, when the quark and antiquark are nearby, the string tension is larger than two fundamental strings and is compatible with the Casimir scaling.
}

\FullConference{The XXV International Symposium on Lattice Field Theory\\
                 July 30-4 August 2007\\
                 Regensburg, Germany}

\begin{document}

\section{Introduction and Motivation}

\begin{figure}
\centering
\includegraphics[width=0.4\textwidth]{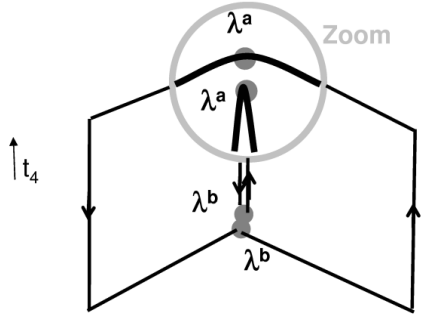}
\caption{Wilson loop for the $q{\overline q}g$ potential,
and equivalent position of the static antiquark, gluon, and quark.} 
\label{Wilson Orlando}
\end{figure}

Our aim is to compute the static potential for the hybrid quark-antiquark-gluon system. In principle any Wilson loop with a geometry similar to the one in Fig.  \ref{Wilson Orlando} and describing correctly the quantum numbers of the hybrid is adequate, although the signal to noise ratio may depend in the choice of the Wilson loop. A correct Wilson loop must include a SU(3) octet, the gluon, a SU(3) triplet, the quark and a SU(3) anti-triplet, the antiquark. It must also include the connection between the three links of the gluon, the quark and the antiquark.

\begin{figure}
\centering
\includegraphics[width=0.4\textwidth]{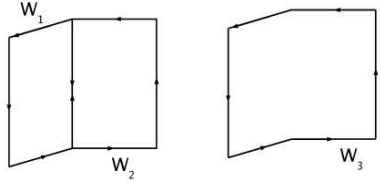}
\caption{Simple Wilson loops that make the qqg Wilson loop.} 
\label{Wloops}
\end{figure}

In the limit of infinite quark mass, a nonrelativistic potential $V$ can be derived from the large time behaviour of euclidean time propagators. Typically, one has a meson operator $\mathcal{O}$ and computes the Green function,
\begin{equation}
   \langle 0| \, \mathcal{O} (t) \, \mathcal{O} (0) \, | 0 \rangle ~
  \longrightarrow ~  \exp \{ - V t \}
\label{Green function}
\end{equation}
for large $t$. Different types of operators allow the definition of different potentials. In the static 
gluon-quark-antiquark interaction, the static gluon can be replaced by a static quark-antiquark pair in a colour octet representation. In this way, we can construct the gluon-quark-antiquark Wilson loop starting from the mesonic operator, 
\begin{equation}
\mathcal{O} (x) ~ = ~ \frac{1}{4} \, 
       \Big[ {\overline q} (x) \,  \lambda^a \, \Gamma_1 \, q(x) \Big]
       \Big[ {\overline q} (x) \,  \lambda^a \, \Gamma_2 \, q(x) \Big]
  \, ,
\label{mesonmeson}
\end{equation} 
where $\Gamma_i$ are spinor matrices. Using the lattice links to comply with gauge invariance, the second operator in eq. (\ref{mesonmeson}) can be made nonlocal to separate the quark and the antiquark from the gluon,
\begin{eqnarray}
 \mathcal{O} (x) & = & \frac{1}{4}
  \Big[ {\overline q} (x) \,  \lambda^a \, \Gamma_1 \, q(x) \Big]
\nonumber \\
   &  & 
 \Big[ {\overline q}(x - r_1 \hat{\mu}_1) 
             U_{\mu_1} (x - r_1 \hat{\mu}_1) \cdots U_{\mu_1} (x - \hat{\mu}_1)
\nonumber \\
   &  &   \lambda^a ~ \Gamma_2  \ \  U_{\mu_2} (x) \cdots U_{\mu_2} (x + (r_2-1)\hat{\mu}_2)
\nonumber \\
   &  &  
         q (x + r_2 \hat{\mu}_2 ) \Big] \, .
\label{op88}
\end{eqnarray}
The nonrelativistic potential requires the computation of the Green functions present in 
eq. (\ref{Green function}). Assuming that all quarks are of different nature, the contraction of the quark field operators gives rise to the gluon operator,
\begin{eqnarray}
W_{O}= &&
\frac{1}{16} 
\mbox{Tr} \Big\{ 
  U^\dagger_4 (t-1,x) \cdots U^\dagger_4 (0,x)  ~  \lambda^b 
\nonumber \\
  &&
   U_4 (0,x) \cdots U_4 (t-1,x)  ~ \lambda^a \Big\} ~ \times
\nonumber \\
\mbox{Tr} 
\Big\{ &  &
   U_{\mu_2} (t,x) \cdots U_{\mu_2} (t,x+(r_2-1)\hat{\mu}_2) 
\nonumber \\
  &&
   U^\dagger_4 (t-1,x + r_2 \hat{\mu}_2) \cdots 
                 U^\dagger_4 (0,x+r_2\hat{\mu}_2) 
\nonumber \\
  &&
   U^\dagger_{\mu_2} (0,x + (r_2-1) \hat{\mu}_2) \cdots 
                 U^\dagger_{\mu_2} (0,x) ~  \lambda^b 
\nonumber \\
  & &
   U^\dagger_{\mu_1} (0,x - \hat{\mu}_1) \cdots 
                 U^\dagger_{\mu_1} (0,x - r_1 \hat{\mu}_1 )
\nonumber \\
  &&
   U_4 (0,x - r_1 \hat{\mu}_1) \cdots 
                 U_4 (t-1,x-r_1\hat{\mu}_1)
\nonumber \\
  &&
   U_{\mu_1} (t,x - r_1 \hat{\mu}_1) \cdots 
                 U_{\mu_1} (t,x - \hat{\mu}_1 ) ~ \lambda^a \Big\} ~ .
\label{glue88}
\end{eqnarray}
Using the relation
\begin{equation}
\sum_a \, \left( \frac{\lambda^a}{2} \right)_{ij} \, 
          \left( \frac{\lambda^a}{2} \right)_{kl} ~ = ~
    \frac{1}{2} \delta_{il} \delta_{jk} - 
     \frac{1}{6} \delta_{ij} \delta_{kl} \, .
\label{exchange+identity}
\end{equation}
one can prove that
\begin{equation}
W_O = W_1 W_2 - \frac{1}{3} W_3
\end{equation}
where $W_1$, $W_2$ and $W_3$ are the simple Wilson loops shown in \ref{Wloops}. This operator is gauge invariant, it is a function of gauge invariant operators (the mesonic Wilson loops). In particular,
for $r_1 = 0$, $W_1 = 1$ and $W_2 = W_3$, the operator reduces to the mesonic Wilson loop. In the
other hand, when $ \hat{\mu} = \hat{\nu} $ and $r_1 = r_2 = r$, $W_2 = W_1^{\dagger}$ and $W_3 = 1$, so $W_O$ reduces to
\begin{equation}
W_O (r,r,t) = |W(r,t)|^2 - 1
\end{equation}
that is the Wilson loop in the adjoint representation \cite{bali} used in the computation of  the potential between two static gluons.

\section{Details of the Simulation}

Our simulation uses the pure gauge $SU(3)$ Wilson action. For a lattice $24^3 \times 48$ and 
$\beta = 6.2$, 141 configurations  were generated, via a combination of Cabbibo-Mariani and overrelaxed updates, with the version 6 of the MILC code \cite{MILC}.

In order to increase the ground state overlap and to reduce the statistical error, prior to the calculation of the operator we applied smearing techniques. In each smearing iteration, the link was replaced
according to
\begin{equation}
U_{\mu}(s) \rightarrow P_{SU(3)} \frac{1}{1 + 6 w} \big(
	U_{\mu}(s) + w \sum_{\mu \neq \nu} U_{\nu}(s) U_{\mu}(s+\hat{\nu})  U_{\nu}^{\dagger} ( s + \hat{\mu} )
	\big)   ;
\end{equation}
we used 25 iterations of smearing in the three spatial directions, with $w = 0.2$, and one iteration in the time direction with $w = 1.0$. Note that while the smearing in the spatial directions only changes the contribution of the various states, the smearing in the time direction also changes the potential, namely it affects the short range part of the potential while leaving the long range behaviour untouched. Nonetheless, the time smearing has the effect of reducing the statistical errors while still getting the right continuous limit (as long as the number of iterations remain small).

\section{The quark-antiquark-gluon Geometries}

In this work only two geometries for the hybrid system were investigated: the perpendicular geometry, where the position of the quark and the position of the antiquark with respect to the gluon form a right angle, and the parallel geometry where this angle is null.

\section{Getting the potential}

\begin{figure}
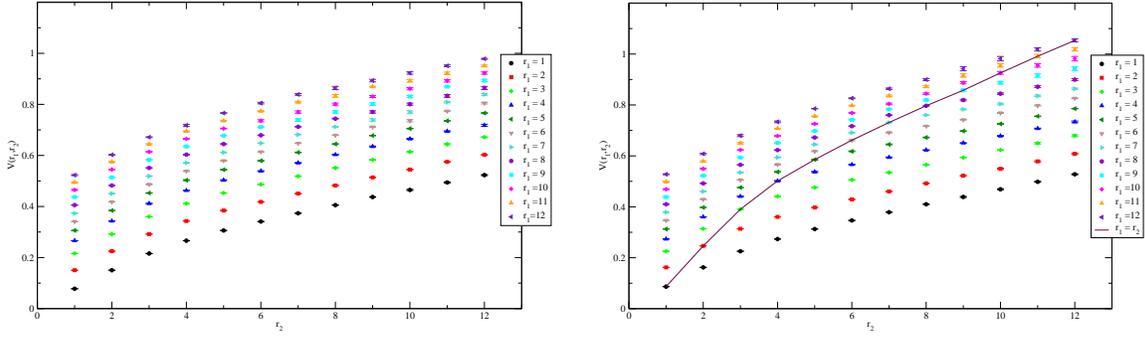

\centering
\includegraphics[width=0.48\textwidth]{Vperp.eps}
\hfill
\includegraphics[width=0.48\textwidth]{Vpara.eps}
\caption{ Graphic of the potential for the perpendicular geometry ( left ) and the parallel geometry ( right )
 as a function of $r_1$ and $r_2$.  } 
\label{potentials}
\end{figure}

The static potential $V_0$ is computed from the large time behaviour of  the $qqg$ Wilson loop,
\begin{equation}
W_O(r_1,r_2,t) = \sum_n C_n(r_1,r_2) e^{ -V_n t }  \longrightarrow C_0(r_1,r_2) e^{ -V_0 t }
\hspace{0,5cm} t \gg 1. 
\end{equation}
To measure $V_0$,  $ - log W(r_1,r_2,t) $ is fitted to a single exponential, choosing a fitting range 
$[t_{min}, t_{max}]$ such  $\chi^2/dof \sim 1$. The potential for the two geometries are reported in
Fig.  \ref{potentials}.

\begin{figure}
\centering
\includegraphics[width=0.48\textwidth]{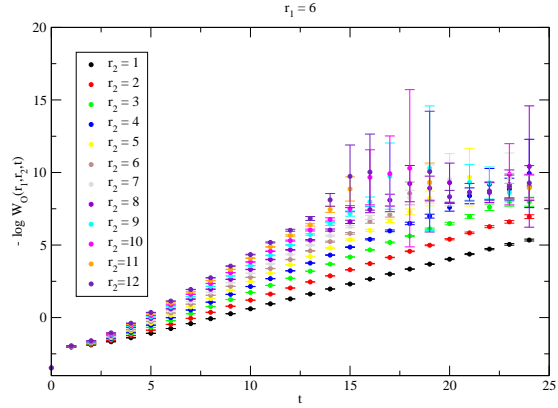}
\caption{ plot of $-log W_O$ as a function of $r_2$ and $t$ for $r_1 = 6$.  } 
\label{logW}
\end{figure}

\section{Results}

\begin{figure}
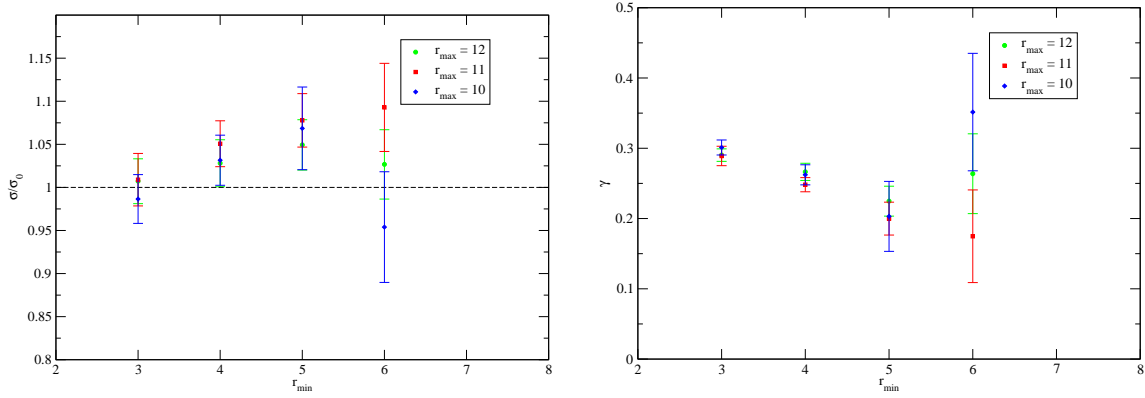

\centering
\includegraphics[width=0.48\textwidth]{sigma_perp.eps}
\hfill
\includegraphics[width=0.48\textwidth]{coulomb_perp.eps}
\caption{ Fitted values of $\sigma$ and $\gamma$ for different values of $r_{min}$ and $r_{max}$ ( perpendicular geometry ) }
\label{resperp}
\end{figure}

\begin{figure}
\centering
\includegraphics[width=0.48\textwidth]{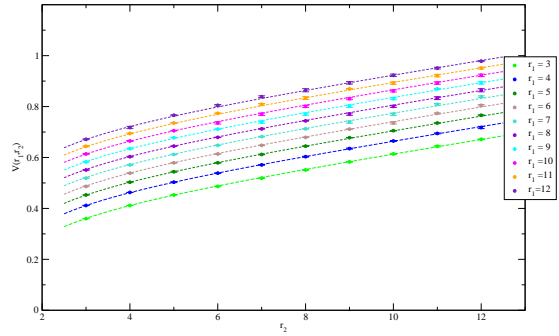}
\caption{ Results for the perpendicular potential and best fit for $r_{min} = 3$ and $r_{max} = 12$ }
\label{perpfit}
\end{figure}

For the perpendicular geometry, the the $qqg$ potential is well described by the ansatz
\begin{equation}
V_{\perp}(r_1,r_2) = c + \sigma ( r_1 + r_2 ) - \gamma ( \frac{1}{r_1} + \frac{1}{r_2} ) ,
\label{VperpFit}
\end{equation}
which assumes that the quark and the antiquark are both linked to the gluon by two independent strings of the same kind. So far we have not considered the possibility of a direct iteraction between the quark and the antiquark. The data points used to fit (\ref{VperpFit}) are choosen in such a way that both $r_1$ and $r_2$ are not larger than a given choosen radius $r_{max}$ and are not smaller than another choosen radius $r_{min}$. These radius were choosen to suppress the finite lattice spacing (volume and noise) effects in the points with smaller (larger) $r_1$ ($r_2$). The results are shown in Fig. \ref{resperp} and \ref{perpfit}.  The data is consistent with $\sigma = \sigma_0$, where $\sqrt{\sigma_0} = 440$ MeV
is the singlet string tension, and $\gamma = \pi / 12$ is the L\"uscher term. 

For the parallel geometry, we start discussing the special case $r_1 = r_2 = r$, which, as already mentioned, corresponds to the adjoint Wilson loop. This system was already studied in \cite{bali} where 
the measured string tension was shown to be consistent with Casimir scaling.

\begin{figure}
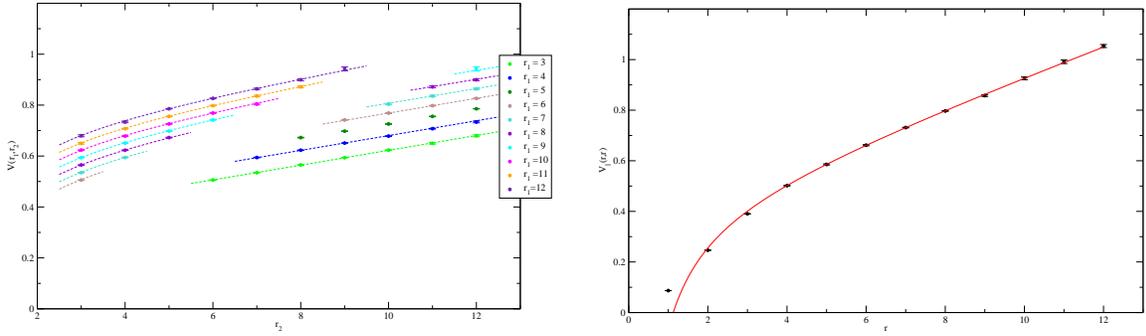

\centering
\includegraphics[width=0.48\textwidth]{parafit.eps}
\hfill
\includegraphics[width=0.48\textwidth]{fitrr.eps}
\caption{ Results for the parallel potential and best fit for $r_{min} = 3$ and $r_{max} = 12$
and points and fit for the parallel potential when $r_1 = r_2$ ( $r_{min} = 4$ and $r_{max} = 12$ )
}
\label{parafit}
\end{figure}

The lattice data was fitted to
\begin{equation}
V(r) = c - \frac{\gamma}{r} + \sigma r .
\end{equation}
For $r_{min} = 4$ and $r_{max} = 12$, we obtained
\begin{equation}
\frac{\sigma}{\sigma_0} = 2.21 \pm 0.06 \hspace{0.75cm}  ( \chi^2/dof = 0.47),
\end{equation}
in agreement with the Casimir scaling hipothesis, i.e. $\sigma = \frac{9}{4} \sigma_0$ (see Fig. \ref{parafit}), and with the results of previous investigation \cite{bali}.

\begin{figure}
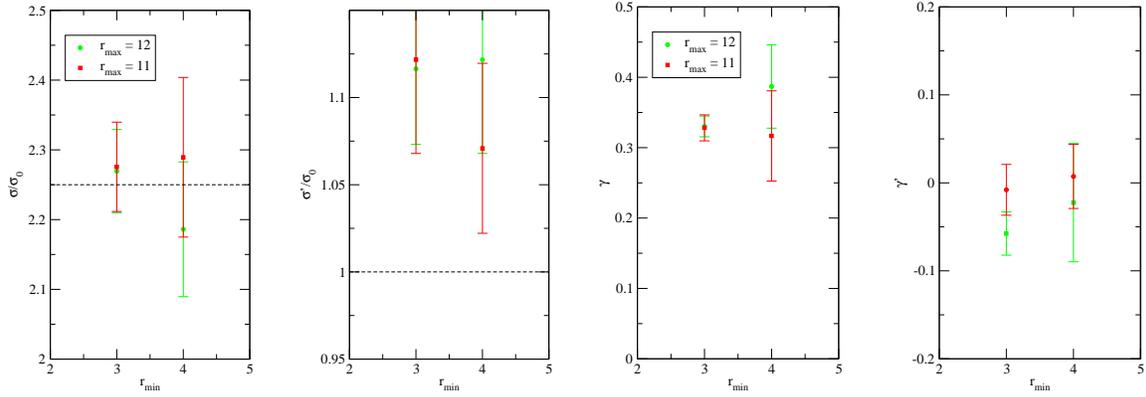

\centering
\includegraphics[width=0.48\textwidth]{sigma_para.eps}
\hfill
\includegraphics[width=0.48\textwidth]{coulomb_para.eps}
\caption{ Fitted values of $\sigma$, $\sigma'$, $\gamma$ and $\gamma'$ for different values of $r_{min}$ and $r_{max}$
( parallel geometry )
} 
\label{respara}
\end{figure}

For the parallel geometry with $r_1 \ne r_2$, the potential was fitted to
\begin{equation}
V_{\parallel} ( r_1, r_2 ) = c + \sigma r_{near} - \frac{\gamma}{r_{near}} + \sigma' | r_2 - r_1 | - \frac{\gamma \, '}{| r_2 - r_1 |}
\end{equation}
where $r_{near} = \min( r_1 , r_2 )$. This expression assumes that the gluon is linked only to the nearest quark by a string, while the quark and the antiquark are linked by a different type of string. The lattice data was chosen as in the perpendicular geometry and requiring 
also that $ | r_1 - r_2 | \leq r_{min} $. The fits are reported in Figs. \ref{respara} and \ref{parafit}. Note that
the string tension associated with the gluon, $\sigma$, agrees with Casimir scaling, i.e.
$\sigma  / \sigma_0 = 9 / 4$, while the quark-antiquark string tension $\sigma'$ agrees with
$\sigma' = \sigma_0$ only at two standard deviations. In what concerns the Coulombic interaction,  
$\gamma$ is larger than the Luscher term but smaller than  $9/4$ of this value. The quark-antiquark
Coulombic coefficient $\gamma \, '$ is consistent with zero.

\section{Results and Conclusions}

In this work we report on the computation of the static potential for the quark-antiquark-gluon system. The details of computation can be found in \cite{qqg}. Two geometries were considered. For the
geometries discussed, the lattice data for the static potential is well described by a sum of two funnel potentials. Furthermore, we confirm Casimir scaling for the adjoint representation.

For the perpendicular geometry, the lattice data supports a gluon linked with the quark and the antiquark by two fundamental strings. For the parallel geometry the string is no longer a fundamental string. Probably, this is due to the repulsion of the strings when they overlap.  For the parallel geometry, despite our data being well described by the ansatz discussed previously, one should try other models. Remember that the Coulomb terms are not consistent with Casimir scaling. We are currently investigating the possibility of linking both the quark and the antiquark directly to gluon and possible contributions from multiple funnel potentials. We also aim to investigate other geometries, namely
the antiparallel one where the position of the quark and the antiquark forms a straight angle.

\end{document}